\documentclass[]{aastex631}
\usepackage{amsmath}
\usepackage{CJK}

\shorttitle{Kinetic equilibrium model of flux ropes}
\shortauthors{Yang et al.}
\graphicspath{{./}{figures/}}

\begin{document}
\begin{CJK*}{UTF8}{gbsn}

\title{Kinetic-scale flux ropes: Observations and applications of kinetic equilibrium models
}

\correspondingauthor{Xu-Zhi Zhou}
\email{xuzhi.zhou@gmail.com}

\author[0000-0001-7591-1093]{Fan Yang (杨帆)}
\affiliation{School of Earth and Space Sciences, Peking University, Beijing, China}

\author[0000-0003-4953-1761]{Xu-Zhi Zhou (周煦之)}
\affiliation{School of Earth and Space Sciences, Peking University, Beijing, China}

\author[0000-0002-5579-005X]{Jing-Huan Li (李京寰)}
\affiliation{School of Earth and Space Sciences, Peking University, Beijing, China}

\author[0000-0002-6414-3794]{Qiu-Gang Zong (宗秋刚)}
\affiliation{School of Earth and Space Sciences, Peking University, Beijing, China}

\author[0000-0002-6059-2963]{Shu-Tao Yao (姚淑涛)}
\affiliation{Shandong Provincial Key Laboratory of Optical Astronomy and Solar-Terrestrial Environment, Institute of Space Sciences, Shandong University, Weihai, China}

\author[0000-0001-6835-4751]{Quan-Qi Shi (史全岐)}
\affiliation{Shandong Provincial Key Laboratory of Optical Astronomy and Solar-Terrestrial Environment, Institute of Space Sciences, Shandong University, Weihai, China}

\author[0000-0001-8823-4474]{Anton V. Artemyev}
\affiliation{Institute of Geophysics and Planetary Physics, University of California, Los Angeles, Los Angeles, CA, USA}



\begin{abstract}

Magnetic flux ropes with helical field lines and strong core field are ubiquitous structures in space plasmas.
Recently, kinetic-scale flux ropes have been identified by high-resolution observations from Magnetospheric Multiscale (MMS) spacecraft in the magnetosheath, which have drawn a lot of attention because of their non-ideal behavior and internal structures.
Detailed investigation of flux rope structure and dynamics requires development of realistic kinetic models.
In this paper, we generalize an equilibrium model to reconstruct a kinetic-scale flux rope previously reported via MMS observations.
The key features in the magnetic field and electron pitch-angle distribution measurements of all four satellites are simultaneously reproduced in this reconstruction.
Besides validating the model, our results also indicate that the anisotropic features previously attributed to asymmetric magnetic topologies in the magnetosheath can be alternatively explained by the spacecraft motion in the flux rope rest frame.  

\end{abstract}




\section{Introduction} \label{sec:intro}

Magnetic flux ropes are frequently observed magnetic structures with helical magnetic field lines and strong core field. In spacecraft observations, flux ropes are often identified from bipolar variations in one magnetic field component and enhancement of magnetic strength at the center. 
There are various observations of flux ropes in the magnetosphere \citep[e.g.,][]{Khurana1995,Slavin2003,Yang2013,Sun2019,Poh2019}, boundary of magnetosphere \citep[e.g.,][]{Rijnbeek1984,Kawano1997,Fear2008,Fear2009,Akhavan-Tafti2018,Hwang2018,Yao2020}, and solar wind \citep[e.g.,][]{Zheng2018,BlancoCano2019,Bai2020}.
Flux ropes are also observed in magnetospheres of Mercury \citep{Zhong2020a}, Mars \citep{Briggs2011}, Jupiter \citep{Sarkango2021}, and Saturn \citep{Jasinski2016}.
Flux ropes are widely believed to result from magnetic reconnection. In earlier studies, some flux ropes near the magnetopause are referred to as flux transfer events (FTEs) \citep{Russell1978}, which allow the plasma transportation across the magnetopause. Various mechanisms for FTE generation have been proposed, such as single X-line reconnection \citep{Scholer1988,Southwood1988} and multiple X-line reconnection \citep{Lee1985}.
Recent simulations show that reconnection can generate not only fluid-scale flux ropes but also those with ion- or even electron-scales \citep{Daughton2011, Hoilijoki2019, Lu2020}. These kinetic-scale flux ropes may interact with one another to form larger-scale flux ropes \citep{Daughton2011}, which could be manifested in spacecraft observations as entangled flux ropes \citep{Wang2017, Oieroset2019,Qi2020}. There are also observations of ion-scale flux ropes near the ion/electron diffusion region \citep{Wang2016,Hwang2018,Poh2019,Dong2020}.
The flux ropes can change in size while moving through convection, often expanding \citep{Dong2017, Akhavan-Tafti2018,Akhavan-Tafti2019} and sometimes contracting \citep{Hasegawa2016}. \cite{Eastwood2012} investigated that the flux ropes in the dayside and in the distant-tail magnetopause have similar orientations and comparable magnetic flux content, indicating that the flux ropes may be in quasi-equilibrium as they are convected tailwards along the magnetopause. Moreover, ion-scale flux ropes are believed to be responsible for exciting waves \citep{Huang2016,Wang2019} or accelerating particles \citep{Zhu2019,Zhong2020}, implying that they are actively involved in magnetospheric physics.

Observations of flux ropes in the magnetosphere often show a force-free configuration, i.e., the electric current is predominantly magnetic field-aligned \citep[e.g.,][]{Slavin2003, Yang2013}. Some flux ropes are classified as being linear force-free as the current density linearly depends on the field strength. Although observations show that many flux ropes are nonlinear force-free structures \citep{Yang2013}, the linear force-free models are still widely used to match the magnetic field observations of flux ropes for simplicity \citep{Akhavan-Tafti2018,Akhavan-Tafti2019}. 
\cite{Eastwood2016} present a typical ion-scale force-free flux rope, which is supported by electron current and embedded in a steady ion flow. This observation shows current filaments and non-ideal ion behavior. Similar features are also found in various studies \citep{Varsani2014,Wang2017,Yao2020}. On the other hand, there are also observations of non-force-free flux ropes, indicating that they are actively evolving \citep{Zhao2016, Teh2017, Bai2020}. \cite{Sun2019} carried out a statistical study to show that quasi-2D ion-scale flux ropes are more likely to be force-free structures. These flux ropes can often be approximated by models with a circular cross section, with larger core fields and lower plasma pressure near the center.

Observational investigation of kinetic-scale flux ropes requires high spatial- and temporal-resolution measurements of electromagnetic field and particle distribution functions, which become available after the successful launch of the Magnetospheric MultiScale (MMS) mission \citep{Burch2016}. MMS consists of 4 identical spacecraft with an average separation of $\sim$10 km, small enough to use multi-spacecraft observation technique \citep[e.g.,][]{Shi2005,Yao2016,Yao2018} to analyze kinetic-scale structures. The FluxGate Magnetometer \citep[FGM,][]{Russell2016} and Fast Plasma Instrument \citep[FPI,][]{Pollock2016} onboard can give high-resolution measurement of magnetic field and particle distributions. 
One way to better understand these observations is to reconstruct these magnetic structures. There are methods of reconstruction using magnetic field measurements only \citep[e.g.,][]{Fu2015,Guo2016}. There are also methods using electromagnetic field and particle moment measurements, which are developed based on the electron magnetohydrodynamics (EMHD) theory and applied to reconstruct fine structures near the electron diffusion region  \citep{Sonnerup2016, Hasegawa2021}.

To make full use of magnetic field and plasma distribution measurement simultaneously, one can carry out the reconstruction using theoretical kinetic models. There are many attempts of constructing such models for current sheets \citep{Allanson2015,Neukirch2020}, magnetic holes \citep{Shustov2016,Li2020,Li2021}, and flux ropes \citep{Vinogradov2016,Allanson2016,Ng2020}. 
These models are equilibrium solutions of Vlasov-Maxwell equations, which provide the location-dependent electromagnetic field and particle distribution functions self-consistently. Although these models describe different structures, their construction shares a similar methodology: choosing invariants of motion, construct the distribution functions using invariants of motion, and solve for the electromagnetic field (see Section \ref{sec:model} for details). We note that in our recent work upon magnetic hole models \citep{Li2020}, we introduced adiabatic invariant (magnetic moment) into the model to better explain the anisotropic feature of pitch angle distribution. We have also used this technique to reconstruct a helical magnetic cavity (similar to flux ropes but with weak core field) \citep{Li2021}. However, there has been no attempt of combining flux rope model (e.g., \cite{Vinogradov2016}) with a strong core field and adiabatic invariants yet.



In this paper, we will generalize the \cite{Vinogradov2016} model to reconstruct the \cite{Yao2020} observations of a kinetic-scale flux rope in the magnetosheath (i.e., the shocked solar wind). In Section \ref{sec:obs}, we revisit the \cite{Yao2020} observations and show their interpretation. In Section \ref{sec:model}, we will introduce the \cite{Vinogradov2016} model in detail and then generalize the model. In Section \ref{sec:recon}, we will introduce how we use the model to reconstruct the flux rope observation and give our interpretation of the observations. In Section \ref{sec:sum}, we conclude the paper with a brief discussion on our interpretations alternative to \cite{Yao2020}. 

\section{Observation Revisited}\label{sec:obs}

\begin{figure}[htb!]
\plotone{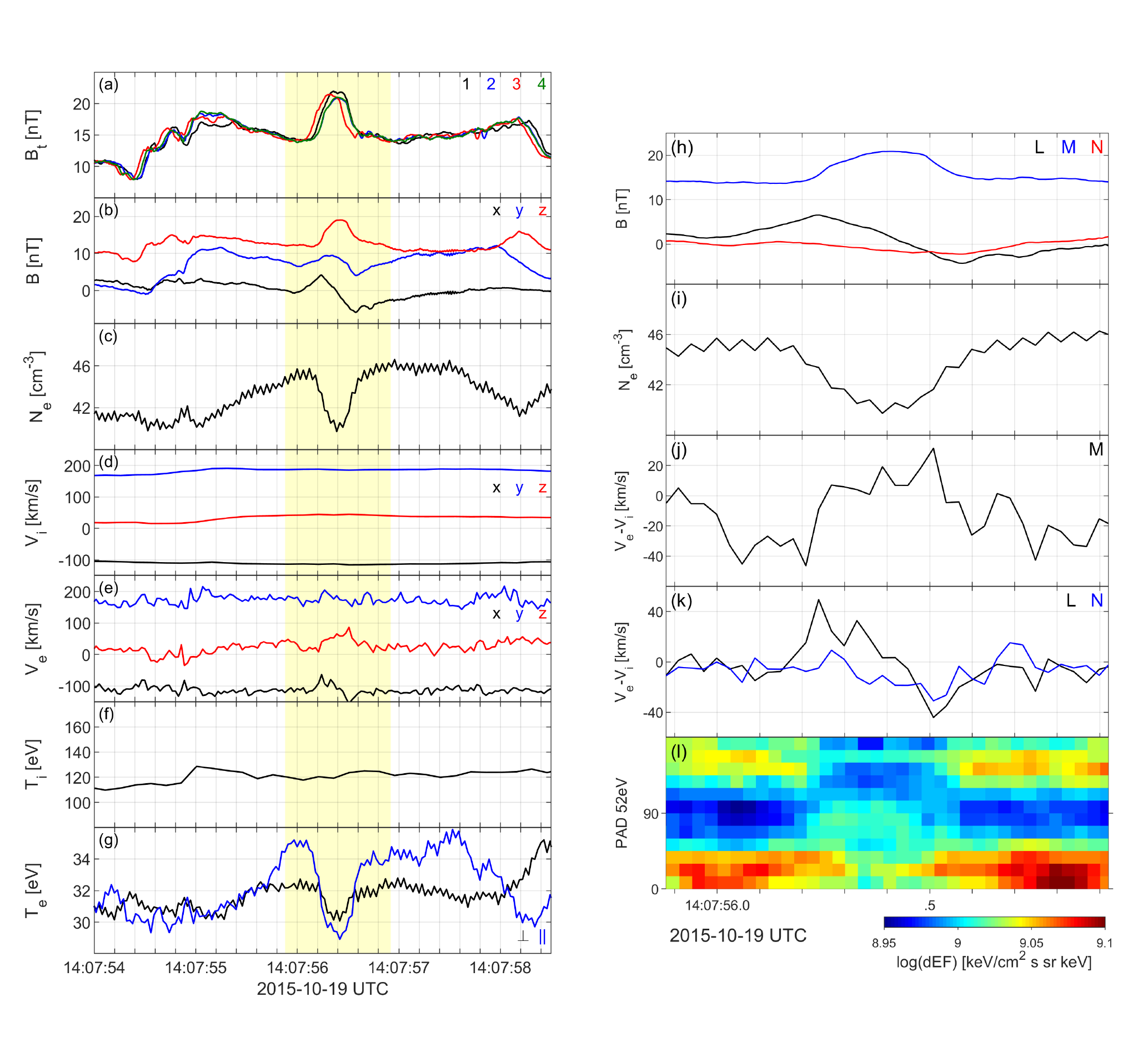}
\caption{Overview of the flux rope observations, with the right panels showing the small-scale observations that correspond to the shaded time interval in the left panels. (a) Magnetic strength by all four MMS
spacecraft. The following panels are observations by MMS1 only. (b)
Magnetic components in GSE coordinates. (c) Electron number
density. (d-e) Ion and electron bulk velocity in GSE coordinates. (f-g)
Ion and electron temperature. In the right panels, the local LMN coordinates are used. (h) Magnetic components. (i) Electron number density. (j-k) Electron bulk
velocity relative to the ion fluid. (l) Electron pitch angle distribution (PAD) at
52 eV, derived from FPI distributions measurement. \label{fig:overview}}
\end{figure}

First, we revisit the kinetic flux rope presented in \citep{Yao2020}.
At $\sim$14:07:56 UT, October 19, 2015, the MMS constellation was inside the magnetosheath, and observed a kinetic-scale flux
rope at {[}8.0, 8.4, -0.8{]} \(R_E\)
(in Geocentric Solar Ecliptic coordinates, or GSE). We replot the magnetic field and plasma property of
this flux rope in Figure \ref{fig:overview}. The four satellites all experienced a
magnetic strength enhancement from 15 nT (the background field) to 22 nT at the flux rope center within a $\sim$0.5s interval (Figure \ref{fig:overview}a), together with a bipolar signature in GSE-x
component of the magnetic field (Figure \ref{fig:overview}b) and a small decrease in plasma number density
(Figure \ref{fig:overview}c). The ion bulk velocity (Figure \ref{fig:overview}d) is less variant than the electron bulk velocity (Figure \ref{fig:overview}e).
The ion and electron temperature (Figure \ref{fig:overview}f-\ref{fig:overview}g) also show moderate change inside the flux rope.
With the assumption of flux rope moving with ion bulk flow and crossed by spacecraft through the core region, the scale of this flux rope can be estimated by \(V_{i\perp}\cdot \Delta t\),
with \(V_{i\perp}\approx 170 \text{km s}^{-1},\Delta t\approx0.5 \text{s}\) being perpendicular ion bulk velocity and duration. 
This estimated scale is comparable to the gyro-radius of a thermal ion \(\rho_i\sim 85\)km, demonstrating the observations of an ion-scale flux rope.

For further analysis, one can use the magnetic field data to perform a
multi-spacecraft analysis \cite[e.g.,][]{Shi2005,Shi2006,zhou2006} to obtain a characteristic orthogonal coordinate system. In \cite{Yao2020}, the minimum directional derivative (MDD) method \citep{Shi2005} was used to define the LMN coordinate system, in which the \(L=[0.96, 0.22, -0.05]\), \(M=[-0.06, 0.46, 0.88]\), and \(N=[-0.22, 0.84, -0.49]\) directions are the maximum, intermediate, and minimum variance directions, respectively. Here we adopt the same coordinate system as in \cite{Yao2020}, and the resulting \(B_M\) and \(B_L\) variations show clear unipolar and bipolar profiles, respectively 
(Figure \ref{fig:overview}h), which demonstrates that the M-direction is along the main axis of the flux
rope. We also present the electron number density (Figure \ref{fig:overview}i) and the electron bulk velocity relative to the ion fluid
(Figure \ref{fig:overview}j-\ref{fig:overview}k), in which the M/L components also show quasi-unipolar/bipolar
signatures. Note that the relative velocity \((V_e-V_i)_M\) is nonzero outside the flux rope, which could result from an offset in association with instrument energy range or photoelectrons, as discussed in \cite{Yao2020}. 
Finally, the electron pitch-angle distributions (PADs) at a typical energy (52eV, between 1-2\(T_e\))
are anisotropic outside the flux rope with higher fluxes in the antiparallel direction than in the parallel and, more significantly, the perpendicular directions (Figure \ref{fig:overview}l). Inside the flux rope, the PADs show a relative concentration of perpendicular-moving electrons. The feature is consistent with the electron temperature profile, in which the parallel temperature is higher than the perpendicular temperature outside the flux
rope, but lower inside. \cite{Yao2020} attributed the anisotropy to the magnetic topology of magnetosheath boundary layer, with the magnetic field lines connecting to the magnetosphere at one end and the magnetosheath at the other. In the following part, we'll bring up a different explanation that the anisotropy is simply because of the background flow. This explanation will be discussed after we develop an equilibrium flux rope model and apply it to the observations.

\section{Model description}\label{sec:model}

As mentioned before, our model is a generalization of a previous
kinetic-scale flux rope model developed by \cite{Vinogradov2016}. The \cite{Vinogradov2016} model is a
self-consistent kinetic equilibrium built in cylindrical coordinates
\((\rho, \varphi, z)\), and all quantities including electromagnetic
fields, particle distribution functions and their moments are functions of \(\rho\) only. In such a system, the charged particles have several
invariants of motion listed as follows:

\begin{equation}
\begin{array}{l}
H_\alpha = \frac{1}{2}m_\alpha v^2 +q_\alpha\phi,\\
P_{\varphi\alpha} = \rho(m_\alpha v_\varphi + q_\alpha A_\varphi),\\
P_{z\alpha} = m_\alpha v_z + q_\alpha A_z,
\end{array} \label{eqn:invariant}
\end{equation}
where the index \(\alpha = i, e\) indicates the particle species (ion or electron). These three invariants are the total energy and two generalized momenta, which are conserved because the system is independent of time and \(\varphi, z\), respectively.
Note that these invariants are functions of electric potential \(\phi\) and magnetic vector potential \(\boldsymbol{A}\). To have a kinetic equilibrium, one may start from constructing the particle distribution functions as functions of these invariants only, which automatically satisfies the equilibrium Vlasov equation. In the \cite{Vinogradov2016} model,
the distribution functions are specified as:

\begin{equation}
\begin{array}{lll}
f_i =& n_i(\frac{m_i}{2\pi T_i})^{3/2}\exp(-\frac{H_i}{T_i}),\\
f_e =& n_e((1+\alpha_r-\alpha_z)(\frac{m_e}{2\pi T_b})^{3/2}\exp(-\frac{H_e}{T_b})\\
& -\alpha_r(\frac{m_e}{2\pi T_r})^{3/2}\exp(-\frac{H_e}{T_r}+\frac{\Omega P_{\varphi e}}{T_r})\\
& +\alpha_z(\frac{m_e}{2\pi T_z})^{3/2}\exp(-\frac{H_e}{T_z}+\frac{VP_{z e}}{T_z})
),
\end{array} \label{eqn:distribution}
\end{equation}
where quantities other than the three invariants are constant parameters. The
ion distribution function is simply Maxwellian (i.e., ions serve as a background carrying no currents), while the electron distribution can be viewed as a linear combination of shifted-Maxwellians. In more detail, each of the three terms in the electron distribution can be understood as the product of two factors, which represent the dependence of the electron phase space densities on velocity and position, respectively. For example, the second term can be factorized as:
\(\exp(-\frac{H_e}{T_r}+\frac{\Omega P_{\varphi e}}{T_r})\propto \exp(-\frac{m}{2T}(v_\varphi - \Omega\rho)^2)\cdot \exp(\frac{e\phi}{T}-\frac{e\Omega\rho A_\varphi}{T}+\frac{m\Omega^2\rho^2}{2T})\),
which indicates a Maxwellian distribution shifted in the velocity space by a constant angular velocity \(\Omega\). The three terms in the electron distribution function
are referred to as the background component, the rotation component and the axial-moving component, labeled by subscripts \(b, r, z\), respectively.
Here, \(n_\alpha, T_\alpha\) are the number density and temperature of each component, \(\alpha_r, \alpha_z\) are coefficients controlling the proportion of the three electron components, \(\Omega,V\) represent
the angular velocity and axial velocity of the corresponding component. The rotation component and axial-moving component introduce the azimuthal and axial currents, both of which depend on \(\rho\). The azimuthal current leads to the strong core field, while the axial current introduces the helical configuration of the magnetic field. These two current-carrying components gradually vanish compared to the background component as \(\rho\to \infty\), making the flux rope smoothly transit to
inhomogeneous magnetic field and plasma. To achieve this, there should be \(\Omega>0\), and the rotation component should have a negative coefficient (i.e., \(\alpha_r>0\)) to model a flux rope with stronger core field (different
from helical magnetic cavity, which has a weaker core field \citep{Li2021}). This negative component may cause risk of negative phase space density. To avoid this, \cite{Vinogradov2016} show that the background component should be hotter than the rotation component (\(T_b>T_r\)).

To solve the electromagnetic field of the flux rope, the prescribed particle distributions are then substituted into the Maxwell equations:

\begin{equation}
\begin{array}{l}N_i = N_e,\\
\frac{1}{\rho}\frac{\partial}{\partial\rho}(\rho\frac{\partial}{\partial\rho} A_z)=-\mu_0J_{ze},\\
\frac{\partial}{\partial\rho}(\frac{1}{\rho}\frac{\partial}{\partial\rho}(\rho A_\varphi)) = -\mu_0J_{\varphi e},
\end{array}    \label{eqn:maxwell}
\end{equation}
where \(N,\boldsymbol{J}\) are number and current densities, and can be integrated from
\(f\). Note that quasi-neutrality equation is used as a valid approximation of the Poisson equation. Although there are other kinetic-scale flux rope models that directly use the Poisson function \citep{Allanson2016, Ng2020}, the quasi-neutrality equation is still used for mathematical simplicity, which is well justified for flux rope scales much larger than the Debye length. A series of boundary conditions are established at the flux rope center, including
\(\phi(0)=0, A_z(0)=0,B_z(0)=B_{0z}\). Note that in such a cylindrical
configuration, \(A_\varphi\) and \(B_\varphi\) are automatically zero at the
center. The two density parameters \(n_i\) and \(n_e\) are also
constrained by the boundary conditions \(N_i(0)=N_e(0)\).
After solving these equations with the above boundary conditions, we can obtain the particle distribution function and electromagnetic profile at any position within this flux rope.

The \cite{Vinogradov2016} model can explain many features of the observed flux rope, e.g., the bipolar signature of magnetic field and the strong core field. However, the model cannot properly reproduce the electron temperature anisotropy within the flux rope. To improve this feature, we adopt methods from our previous model on magnetic cavities \citep{Li2020} to
introduce the magnetic moment \(\mu\) (often referred to as the first adiabatic invariant) as an additional invariant of particle motion in this model:

\begin{equation}
\begin{array}{lll}
\mu_e =& \frac{m_ev_\perp^2}{2B_{rc}},\\
f_i =& n_i(\frac{m_i}{2\pi T_i})^{3/2}\exp(-\frac{H_i}{T_i}),\\
f_e =& n_e((1+\alpha_r-\alpha_z)(\frac{m_e}{2\pi T_b})^{3/2}\exp(-\frac{H_e}{T_b}+\frac{b_b\mu_e}{T_b})\\
& -\alpha_r(\frac{m_e}{2\pi T_r})^{3/2}\exp(-\frac{H_e}{T_r}+\frac{\Omega P_{\varphi e}}{T_r}+\frac{b_r\mu_e}{T_r})\\
& +\alpha_z(\frac{m_e}{2\pi T_z})^{3/2}\exp(-\frac{H_e}{T_z}+\frac{VP_{z e}}{T_z}+\frac{b_z\mu_e}{T_z})
),
\end{array} \label{eqn:distribution-mu}
\end{equation}
where \(B_{rc}\) represents the magnetic strength at the electron gyro-center, and
\(v_{\perp}\) is the perpendicular velocity relative to the gyro-center. In the lowest order of
approximation (\(v_\perp^2=v^2 -v_{||}^2 \approx v^2-v_z^2\)), the newly-introduced \(\mu\)-dependence in the electron distribution can modify the perpendicular temperature, while the parallel
temperature remains unchanged. In other words, \(b_b,b_r,b_z\) are
coefficients controlling the anisotropy of each component. If
\(b_b=b_r=b_z=0\), our model will degenerate to the original \cite{Vinogradov2016} model. Note that in our definition of magnetic moment, the magnetic field profile is pre-required. Therefore, we follow the iteration procedure
adopted in \cite{Li2020}. The iteration procedure takes a magnetic field profile (such as the one in the \cite{Vinogradov2016} model) as the input, and returns an output  electromagnetic field by solving equations \ref{eqn:maxwell} and \ref{eqn:distribution-mu}. The iteration converges when there is no significant difference between the input and output field, and we accordingly obtain the self-consistent
electromagnetic field and distribution functions.

\section{Application of the Model}\label{sec:recon}

Next, we use this new model to reconstruct the flux rope in Figure \ref{fig:overview}. The
first step is to determine the model \((\rho, \varphi, z)\) cylindrical
coordinates on the ground of observational coordinates, e.g., the LMN
coordinates. We can conveniently set the model z-axis to be aligned with the M direction. For consistency the flux rope is assumed to move with the ion bulk flow, since in the model the ion has no bulk velocity. 
However, since the LMN coordinates in \cite{Yao2020} only
give the directions, the center (or the location of the z-axis, projected to the LN plane) in the model coordinates should also be
determined.


We next determine, based on the 1D cylindrical symmetry, the center of the flux rope in its rest frame (or equivalently the rest frame of the ion fluid). In our model, the magnetic field strength only depends on the radial distance \(\rho\). Therefore, the contours of magnetic field strength should form a series of concentric circles in the LN plane. In this frame of reference, each spacecraft moves along a straight line with a constant velocity, i.e., the reverse of the ion bulk velocity in the spacecraft frame (which is nearly constant in reality, see Figure \ref{fig:overview}d). For any given contour of the field strength, one can determine the time and location of intersection (as the observed magnetic field reach the given field strength). Since each spacecraft can traverse each contour twice (once inward and once outward), there would be 8 intersection points for each contour. If the flux rope is ideally 1D cylindrical symmetric, there should be a center point that for each set of intersections, the distances from this center point to the 8 intersections are the same. 
Based on this principle, we adopt a best-fit procedure to minimize the average variance of each set of distances:

\begin{equation}
    L = \frac{1}{N}\sum_i\text{var}_j(d_{ij})=\frac{1}{8N}\sum_i\sum_j (d_{ij} - \text{avg}_j(d_{ij}))^2, \,j=1,2,...,8,
    \label{eqn:cost}
\end{equation}
where \(d_{ij}\) represents the radial distance of the j-th point to the i-th contour, and the undetermined center coordinates are implicitly contained therein. Here, N represents the number of chosen contour lines in this minimization procedure. The \(\text{var}_j\) and \(\text{avg}_j\) functions are the variance and the mean value of \(d_{ij}\) among subscript j, respectively. The best-fit flux rope center is considered as the location that minimize the loss function L,
and the minimization procedure can be performed with gradient descent or any other non-linear optimization algorithm. 

The application of the best-fit procedure is visualized in Figure \ref{fig:center}. We first select 11 values of magnetic field strength ranging from 16 nT to 20 nT, to identify the specific moments in time when the observed field strength matches the given values (see Figure \ref{fig:center}a). The corresponding spacecraft locations of intersection are given in Figure \ref{fig:center}b as the cross symbols, which serve as the input of the minimization procedure that determines the flux-rope center (the asterisk in Figure \ref{fig:center}b) and the series of circular contours. The arrows represent the measured \(B_{LN}\), or the magnetic field component within the LN plane, which are approximately aligned with the local tangents of the circular contours. This feature agrees with expectations, since the radial magnetic field \(B_\rho\) is always zero in the modeled flux rope. We also show in Figure \ref{fig:center}c the scatter plot of \(d_{ij}\) against
\(\text{avg}_j(d_{ij})\), to indicate how well these \(d_{ij}\) values match their mean value \(\text{avg}_j(d_{ij})\), and thus how well each set of intersections
form the concentric circles. The best-fit procedure yields a loss function $L\approx 1.47 \text{ km}^2$, which indicates that the average deviation of the intersections from their corresponding circles equals $\sqrt{L}\approx 1.21\text{ km}$, a number much smaller than the flux rope radius ($\sim 30$ km, see Figure \ref{fig:center}b). A caveat in the best-fit procedure is that the time and location of intersection must be determined accurately. Therefore, we only select the contours with large enough $dB/dt$ values. Given the weaker magnetic gradient near the flux-rope center (see Figure \ref{fig:center}a), this region is avoided in the contour selection, which is also manifested by the central void region without contours in Figure \ref{fig:center}b. 

With the flux-rope center fixed, the cylindrical coordinate system in our model is determined, and we are now ready to apply the kinetic equilibrium model to the observed flux rope. To make sure that our model is consistent with measurements from all the four spacecraft, we adjust the model parameters to match the MMS1 observations, and then examine the consistency between the model and the measurements from the other three spacecraft. The chosen parameters are shown in Table \ref{tab:params}, and the model results are shown in Figure \ref{fig:MMS12} (dashed lines) , Figure \ref{fig:compareJ} (red lines) and Figure \ref{fig:MMS34} (PAD plots in lower panels) along with the spacecraft observations (solid lines in Figure \ref{fig:MMS12} , black and blue lines in Figure \ref{fig:compareJ} and PAD plots in upper panels of Figure \ref{fig:MMS34}). Although there are some disagreement between the model and the observations, the model manage to reproduce most of the key features. 

In Figures \ref{fig:MMS12}a-b, the unipolar \(B_M\) and bipolar \(B_L\) profiles observed by all the four spacecraft can be well reproduced by the model. The \(B_N\) profile also shows a negative peak with magnitude comparable to \(B_L\) except for MMS1, since the MMS1 path is too close to the flux rope center. The modeled number density (Figures \ref{fig:MMS12}c), as expected, shows a local minimum at the flux rope center. We also show in Figures \ref{fig:MMS12}d-e the electron bulk velocity in the ion rest frame. The observed bulk velocity appears to be more variant in general, and in some components there are discernible offsets of background values far away from the flux rope center (e.g., M component in Figure \ref{fig:MMS12}e1,e2 and N component in Figure \ref{fig:MMS12}d3). As mentioned in Section \ref{sec:obs}, the offset probably results from measurement error \citep{Yao2020}, which cannot be reproduced by the model. In spite of this, the measured and modeled velocity differences both show bipolar signatures in the L direction and unipolar signatures in the M direction. The plasma density and electron bulk velocity profiles can be combined to obtain the current density distributions, which are displayed in Figure \ref{fig:compareJ}. The red lines represent the current density in our model, whereas the black lines are the spacecraft measurements. Moreover, the current density can be estimated by calculating the curl of the magnetic field based on the four-spacecraft curlometer method \citep{Dunlop2002}, which is shown as the blue lines in Figure \ref{fig:compareJ}. These modeling and observational results show similar range and variations, which support the self-consistency of our model. We also compare the electron temperature profiles in Figures \ref{fig:MMS12}f-g. Note that we used the diagonal terms of temperature tensor (integrated from distribution function) in the \((\rho,\varphi, z)\) coordinates, whereas the measured parallel and perpendicular electron temperatures are derived from temperature tensor using local magnetic field-aligned coordinates. Since the angle between z-axis and magnetic field direction is very small, we have \(T_{||}\approx T_{zz}, T_{\perp}\approx (T_{\rho\rho} + T_{\varphi\varphi})/2\), and thus the observed temperature and corresponding model results can be compared directly. 

We also present the electron PADs at four different energies (32eV, 40eV,
66eV, 88eV) in Figures \ref{fig:MMS34}b-e and the modeling results in Figures \ref{fig:MMS34}f-i. As mentioned before, the observed PADs display
two main features: (1) the bi-directional flow outside the flux rope, with greater parallel flux than antiparallel flux; (2) the relative concentration of perpendicular flux inside. These two features also appear in the modeling results, although the exact fluxes may not be the same as the observations. Also, we point out that the ratio between the electron fluxes in the parallel and in the antiparallel directions appears to be higher in the modeled flux rope than in the observations. Despite these minor discrepancies, the similar pattern between observations and the modeling results enables us to understand the physics behind these features. The higher fluxes in the parallel direction than in the antiparallel direction originate from the spacecraft motion across the modeled flux rope (i.e., flux rope transport by plasma flow across the spacecraft). In the presence of this spacecraft velocity, the observed electrons with the same energy but moving in different directions would have different energies in the rest frame of the flux rope. Given the Maxwellian (or shifted-Maxwellian) distributions of the electrons in the model, this energy change will exponentially affect the phase space
density. In a previous work \citep{Li2021} this effect is quantified as follows:

\begin{equation}
r_{p/a} = \frac{f_{\text {para }}}{f_{\text {anti }}}=\exp \left(\frac{2 M_{e} v_{s}\left|v_{\zeta}\right|}{T_{e, ||}}\right),
\label{eqn:ratio}
\end{equation}
where \(v_s\) and \(v_\zeta=(2W/M_e)^{1/2}\) represent the parallel component of the spacecraft velocity and the velocity of the electrons at energy \(W\), respectively. 
In our case, although \(v_s\approx 140\) km s$^{-1}$ is far less than the thermal velocity (with \(T_e\approx 33\text{eV}\), \(v_{the}\approx 3400\text{km s}^{-1}\)), this effect is still comparable to the overall electron energy flux variations ($\sim$1.2 times) during this event and therefore cannot be neglected. 
On the other hand, since the electron distribution outside the flux rope is nearly bi-Maxwellian, one can estimate the difference between perpendicular and parallel/antiparallel flux as:
\begin{equation}
r_{\perp/||}=\frac{f_{\perp}}{f_{||}}=\exp \left(\frac{ M_{e} v_{\zeta}^2}{2}\left(\frac{1}{T_{e, ||}} - \frac{1}{T_{e, \perp}}\right)\right).
\label{eqn:ratio_perppara}
\end{equation}
Here the spacecraft velocity $v_s$ is omitted and thus $f_{||}$ is some value between $f_{para}$ and $f_{anti}$.
These two types of anisotropy have different dependence on electron energy, i.e., $\log(r_{p/a})\propto |v_\zeta|$, $\log(r_{\perp/||})\propto v_\zeta^2$.
This is consistent with the first feature both in observations and model results that the ratio between parallel and antiparallel flux dominates at lower energies, while the ratio between perpendicular and parallel flux becomes more significant at higher energies.
The second feature, i.e., the relative concentration of perpendicular flux inside the flux rope, results from proportion changes of different terms in the electron distribution with the distance from the flux rope center.
The current-carrying terms, which have higher
perpendicular temperature than parallel temperature ($b_z > 0, b_r = 0$ in Table \ref{tab:params}), are significant near the center. 
This results in greater perpendicular flux than parallel flux according to Equation \ref{eqn:ratio_perppara}. 
This model can also explain that this
concentration vanishes at higher energy, since the current-carrying terms
have lower temperature than the background temperature, and become less significant at higher energy. 
In short, our reconstruction demonstrates that the
complicated PADs observation can be explained by this simple quasi-equilibrium model.

\section{Summary and discussion}\label{sec:sum}

In this paper, we generalize the self-consistent kinetic model of magnetic flux ropes \citep{Vinogradov2016}, and apply the generalized model to a kinetic-scale flux rope
observed by \cite{Yao2020}. The model can explain most of the
observational features including the magnetic field profiles and the anisotropic electron distributions.

In \cite{Yao2020}, the explanation for the different electron fluxes in the parallel and antiparallel directions is that the flux rope is in the magnetosheath boundary layer,
where the magnetic field line is connected to the open field-line on one side and to the closed field-line on the other. In our
model, however, this feature is caused by the background plasma flow in the axial direction without invoking specific assumptions upon the field-line topology. This may help us understand how this
kind of distribution is formed and whether it's related to larger-scale
magnetospheric structures. This equilibrium model, after validation from observations, can also be used to investigate the stability of magnetic flux ropes. Moreover, this model can provide an appropriate initial condition in kinetic simulations to understand the particle dynamics associated with flux rope evolution.

\begin{figure}[htb!]
\plotone{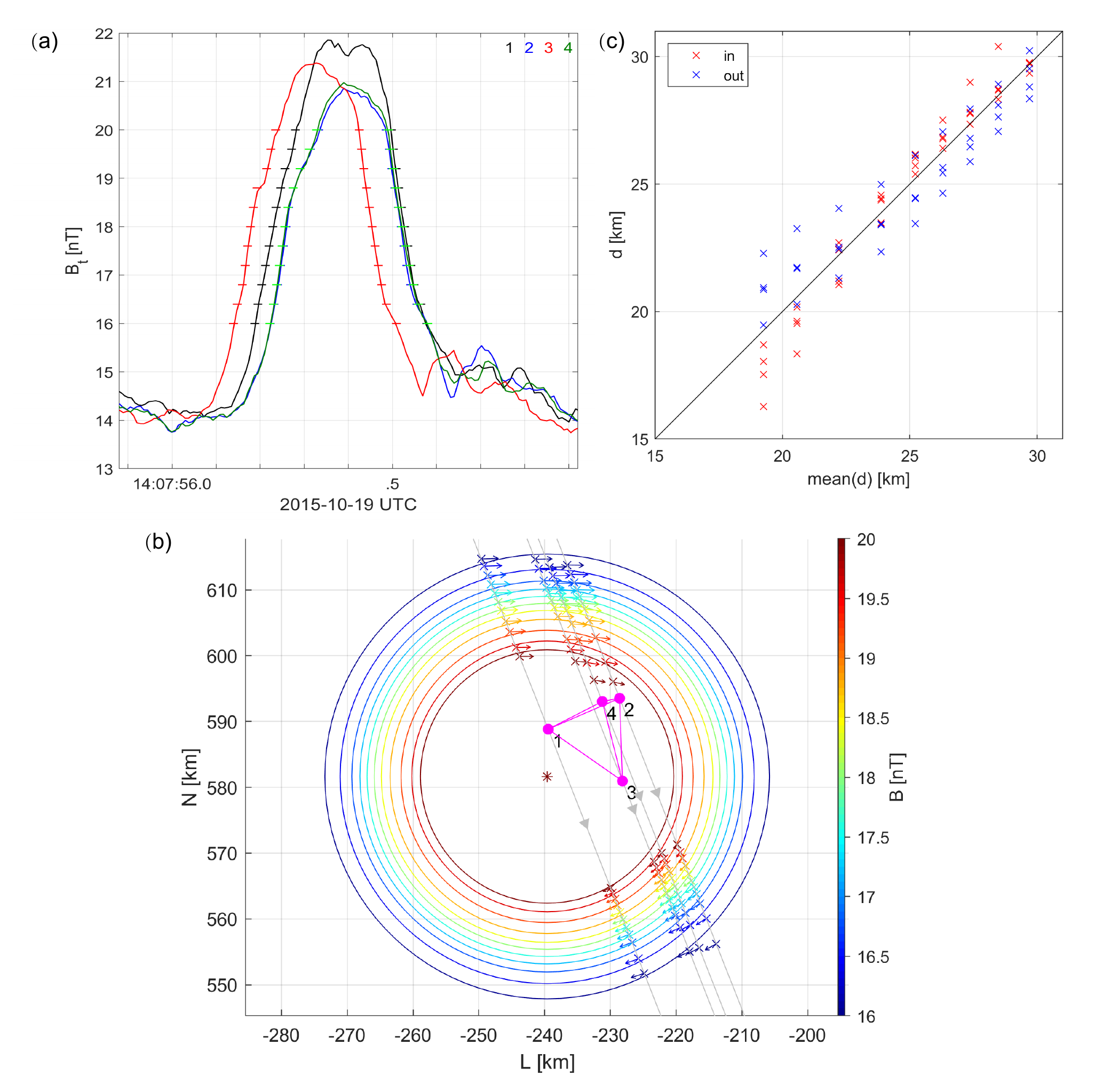}
\caption{Identification of the flux rope center via a best-fit procedure using magnetic field measurements. (a) The magnetic field strength observed by the four MMS spacecraft, in which the short horizontal ticks mark the selected contours of magnetic strength and the identified moments of contour intersection. (b) Visualization of the best-fit procedure in the LN plane. Different color marks the series of contours. The cross symbols mark the intersection location where MMS satellites traverse the magnetic contours (projected to LN plane). The arrows represent the corresponding magnetic components within the LN plane, and the central asterisk marks the fitted flux-rope center. The magenta points marks the projection of spacecraft location into LN plane at a given time (14:07:56.347 UT). (c) Scatter plot showing the radial distance of the intersection points against the average distance for each contour. The inward and outward intersections
of the MMS satellites are marked with red and blue crosses, respectively. The solid black line represents the diagonal line.\label{fig:center}}
\end{figure}

\begin{figure}[htb!]
\plotone{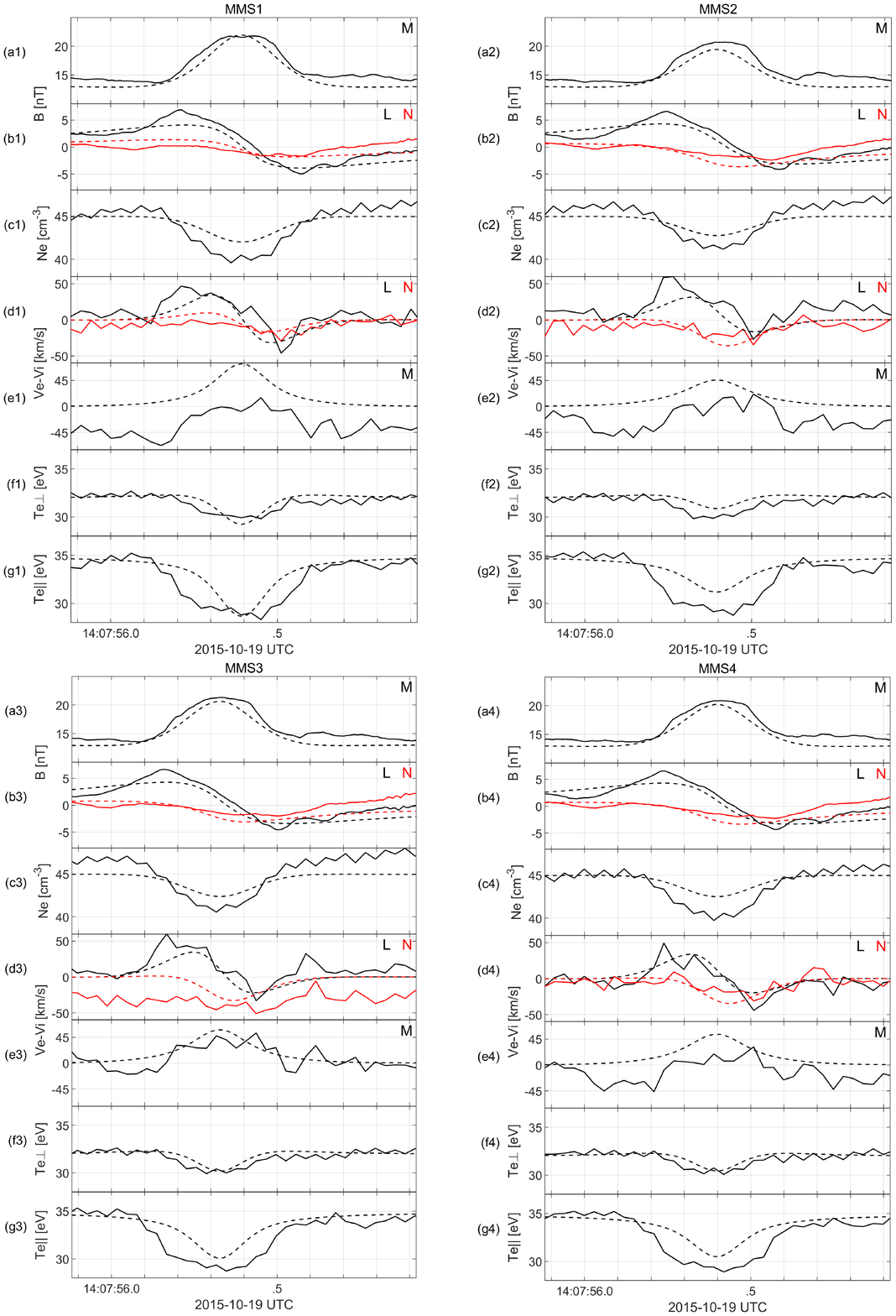}
\caption{Comparison between MMS observations and the modeling results on the magnetic field and electron moments. Each column corresponds to one of the four MMS spacecraft. The solid lines represent the observations, whereas the dashed lines represent the modeling results. (a-b) Magnetic field components in LMN coordinates. (c) Electron Number density. (d-e) Electron bulk velocity in the rest frame of the ion fluid. (f-g) Electron perpendicular and parallel temperature. 
\label{fig:MMS12}}
\end{figure}

\begin{figure}[htb!]
\plotone{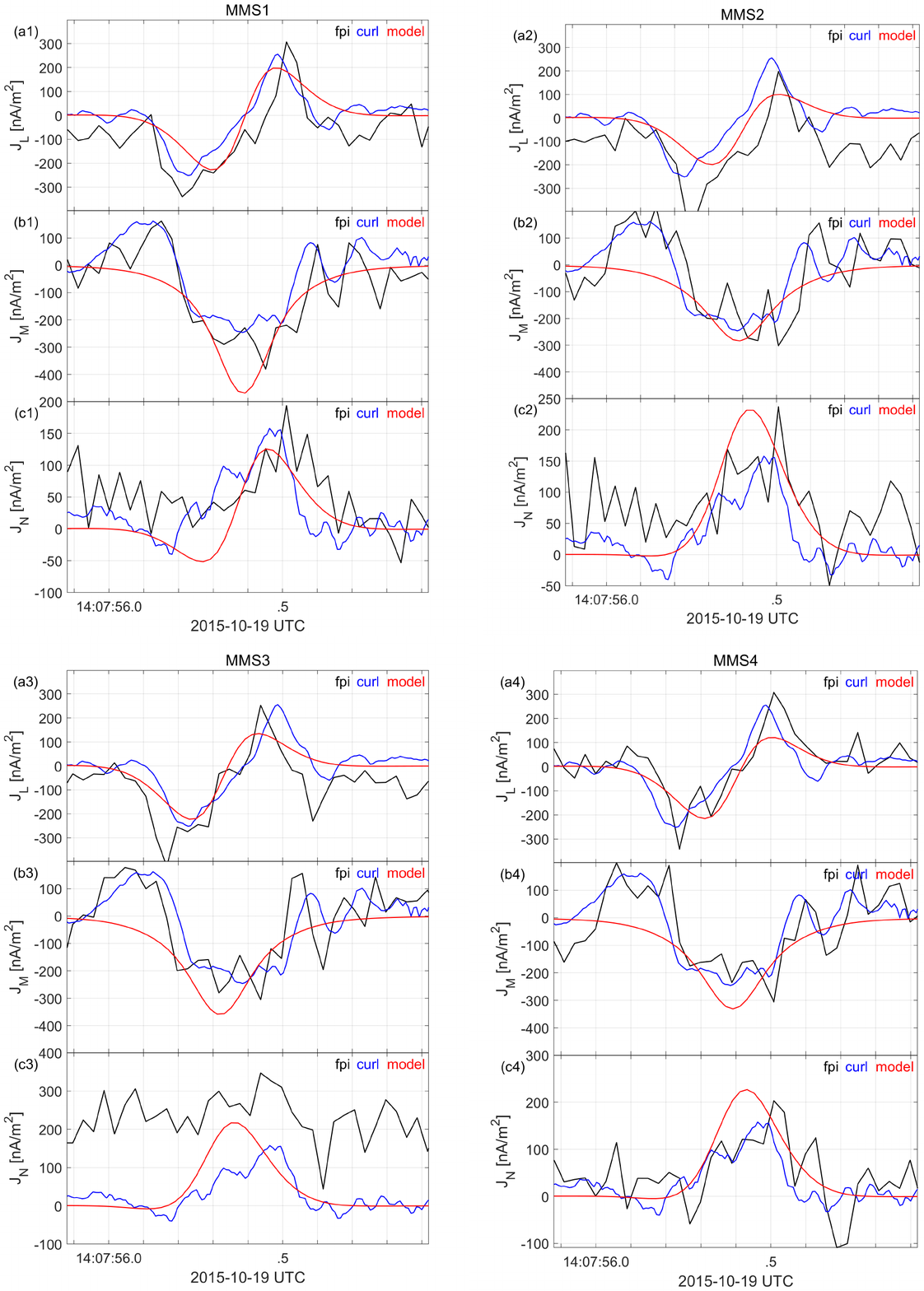}
\caption{Comparison between the current density derived from different methods. Each column corresponds to one of the four MMS spacecraft. (a-c) L, M, and N components of the current density. Black lines represent the current density determined from FPI measurements. Blue lines represent current density derived from multi-spacecraft curlometer method (which are the same for all the four spacecraft). Red lines represent the results from our kinetic model.
\label{fig:compareJ}}
\end{figure}

\begin{figure}[htb!]
\plotone{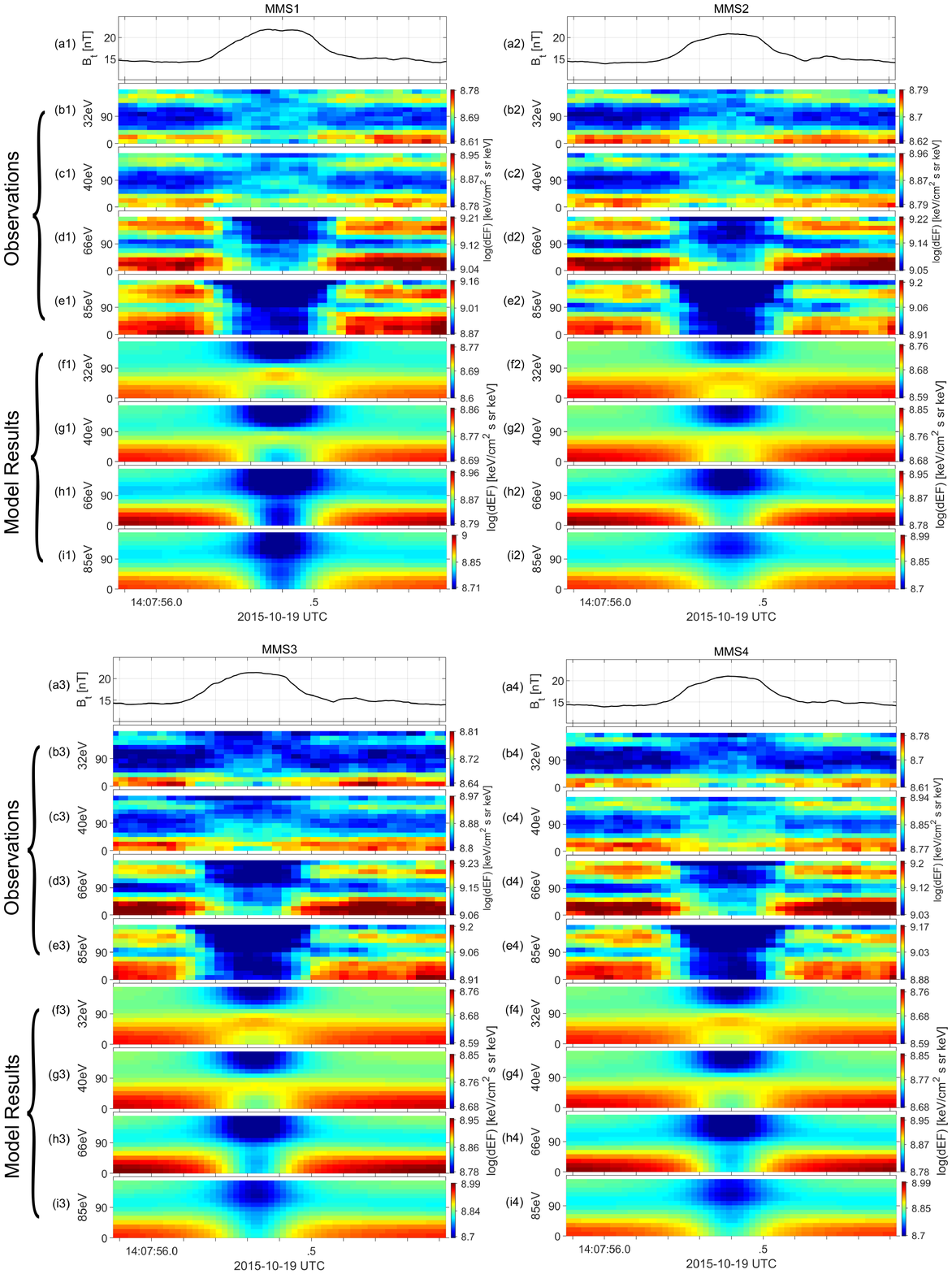}
\caption{Comparison between MMS observations and the modeling result on electron pitch angle distributions. Each column corresponds to one of the four MMS spacecraft. (a) For reference, the observed magnetic field strength. (b-e) The electron pitch angle distributions observed from four consecutive energy channels (32, 40, 66, 85 eV). (f-i) The modeled pitch angle distributions within the four energy channels.
\label{fig:MMS34}}
\end{figure}

Finally, we point out that although our model reproduces many key features of the observed flux rope, their direct comparison in Figures \ref{fig:MMS12}, \ref{fig:compareJ} \& \ref{fig:MMS34} still shows some discernible differences. A most significant disagreement is that the modeled electron fluxes in the antiparallel direction appears to be lower than in the observations. The overestimation of $f_{para}/f_{anti}$, according to Equation \ref{eqn:ratio}, could originate from the error in the determination of the electron temperature or plasma bulk velocity due to the limited energy range of the FPI measurements. It is also possible that the background electron distributions are more complicated than the bi-Maxwellian function adopted in the model. Another source of the disagreement is the error in determining the flux rope center (see Figure \ref{fig:center}), which could indicate a possible deviation of the configuration from cylindrical symmetry. After all, our model is based on a one-dimensional assumption that all quantities only depend on the radial distance from the flux rope center, which is highly idealized and may differ from the observations. The error may also originate from the evolution and/or acceleration of the entire structure. In this case, our equilibrium model can only serve as the initial condition to facilitate the kinetic simulation of flux rope evolution and dynamics.

\begin{acknowledgments}
This study was supported by NSFC grant 42174184. 
We are grateful to the MMS team for providing the high-quality observational data utilized in this study, which are available from the MMS science data center (\url{https://lasp.colorado.edu/mms/sdc/public/}). 
Data analysis was performed using the IRFU-Matlab analysis package available at \url{https://github.com/irfu/irfu-matlab}. 
The model codes are available from the Zenodo repository (\url{https://doi.org/10.5281/zenodo.5555811}).
\end{acknowledgments}


\bibliography{kfr}{}
\bibliographystyle{my_aasjournal}



\begin{deluxetable*}{lll}
\tablenum{1}
\tablecaption{Model parameters used in the reconstruction.}
\tablehead{\colhead{Symbols} & \colhead{Description} & \colhead{Value}}

\startdata
\(\alpha_{r}\) & proportion of negative rotation component &
0.75 \\
\(\alpha_z\) & proportion of axial moving component &
0.35 \\
\(n_i~[\text{cm}^{-3}]\) & ion number density & 42 \\
\(B_{0z} ~[\text{nT}]\) & magnetic strength at center &
22 \\
\(\Omega~[\text{s}^{-1}]\) & angular velocity of rotation component &
5.75 \\
\(V~[\text{km s}^{-1}]\) & axial velocity of axial moving component &
216 \\
\(b_b\) & coefficient of \(\mu\)-term in background component &
-0.07 \\
\(b_r\) & coefficient of \(\mu\)-term in rotation component &
0 \\
\(b_z\) & coefficient of \(\mu\)-term in axial moving component &
0.53 \\
\(T_{eb}~ [\text{eV}]\) & temperature of electron background component &
34.8 \\
\(\tau_r=T_{er}/T_{eb}\) & normalized temperature of electron rotation
component & 1.05 \\
\(\tau_z=T_{ez}/T_{eb}\) & normalized temperature of electron axial
moving component & 0.4 \\
\(\tau_i=T_{i}/T_{eb}\) & normalized temperature of ions &
3.7\label{tab:params}
\enddata
\end{deluxetable*}

\end{CJK*}
\end{document}